\documentclass[aps,prl,twocolumn,superscriptaddress,english]{revtex4-1}
\usepackage[utf8]{inputenc}

\usepackage{amsmath}
\usepackage{amssymb}
\usepackage{graphicx}
\usepackage{xcolor}
\usepackage{lipsum}
\usepackage{xr-hyper}
\usepackage[colorlinks=true,urlcolor=blue,citecolor=blue,linkcolor=blue,breaklinks=true]{hyperref}
\usepackage{lineno}
\usepackage{float}
\usepackage{orcidlink}

\newcommand{\khomit}[1]{}
\makeatletter
\newcommand*{\addFileDependency}[1]{
\typeout{(#1)}
\@addtofilelist{#1}
\IfFileExists{#1}{}{\typeout{No file #1.}}
}\makeatother

\newcommand*{\myexternaldocument}[1]{%
\externaldocument{#1}%
\addFileDependency{#1.tex}%
\addFileDependency{#1.aux}%
}
\myexternaldocument{supplemental}

\newcommand{\Tc}{$T_\text{c}$}
\newcommand{\ep}{$ep$}

\makeatletter
\def\@fnsymbol#1{\ensuremath{\ifcase#1\or \dagger\or *\or \ddagger\or
   \mathsection\or \mathparagraph\or \|\or **\or \dagger\dagger
   \or \ddagger\ddagger \else\@ctrerr\fi}}
\makeatother

\begin{document}

\author{Simone Di Cataldo\,\orcidlink{0000-0002-8902-0125}}\email{simone.cataldo@tuwien.ac.at}
\affiliation{Institut f\"{u}r Festk\"{o}rperphysik, Technische Universit\"{a}t Wien, 1040 Wien, Austria} 
\author{Paul Worm\,\orcidlink{0000-0003-2575-5058}}
\affiliation{Institut f\"{u}r Festk\"{o}rperphysik, Technische Universit\"{a}t Wien, 1040 Wien, Austria} 
\author{Liang Si\,\orcidlink{0000-0003-4709-6882}}
\affiliation{School of Physics, Northwest University, Xi’an 710127, China} 
\affiliation{Institut f\"{u}r Festk\"{o}rperphysik, Technische Universit\"{a}t Wien, 1040 Wien, Austria} 
\author{Karsten Held\,\orcidlink{0000-0001-5984-8549}}
\affiliation{Institut f\"{u}r Festk\"{o}rperphysik, Technische Universit\"{a}t Wien, 1040 Wien, Austria} 

\title{Absence of electron-phonon-mediated superconductivity in hydrogen-intercalated
  nickelates}
\date{\today}

\begin{abstract}
  A recent experiment  [X. Ding {\em et al.}, Nature 615,
    50 (2023)] indicates that superconductivity in nickelates is restricted
  to a narrow window of hydrogen concentration: $0.22<x<0.28$ in Nd$_{0.8}$Sr$_{0.2}$NiO$_2$H$_{x}$. This reported necessity of hydrogen suggests that it  plays a crucial role for superconductivity, as it does in the vast field of hydride superconductors.
Using density-functional theory and its extensions, we explore the  effect of topotactic hydrogen on the electronic structure and phonon-mediated superconductivity in nickelate superconductors.
Our calculations show that the electron-phonon coupling in hydrogen-intercalated nickelates is not strong enough to drive the electron pairing, and thus cannot explain the reported superconductivity.
\end{abstract}

\maketitle

Our understanding of the pairing mechanism and gap function in the recently synthesized nickelate superconductors~\cite{li2019superconductivity,Osada2020,Zeng2021,Osada2021,pan2021,noteTheory}
is still in its infancy, and it goes without saying that it is controversially debated.
Scanning tunneling microscopy (STM) shows both a ``U'' and a ``V'' shape gap~\cite{Gu2020b}, depending on the precise position of the tip on the surface and indicative of a $d$- and $s$-wave gap, respectively. Fits to the London penetration depth either point to a nodeless~\cite{chow2022pairing} or a nodal~\cite{harvey2022evidence} gap.

Theories range from  $d$-wave superconductivity
originating from spin-fluctuations in the Ni $d_{x^2-y^2}$ orbital \cite{Wu2019,Kitatani2020,kitatani2022optimizing,Karp2022} to two-orbital physics with $d$- and $s_{\pm}$-wave superconductivity~\cite{PhysRevLett.129.077002}.
Also superconductivity based on a Kondo coupling between Ni-spin and Nd-bands \cite{Zhang2019}, the importance of the inter-orbital Coulomb interaction~\cite{Adhikary2020},
and a possible connection to 
charge ordering~\cite{rossi2021broken} have been suggested, among others.

Early calculations~\cite{Si2019,NoteLaterH} indicated that topotactic hydrogen might be intercalated  when reducing  Nd$_{0.8}$Sr$_{0.2}$NiO$_3$ to  Nd$_{0.8}$Sr$_{0.2}$NiO$_2$ with  the reagent CaH$_2$~\cite{Lee2020}. The presence of hydrogen in nickelates has by now been established using nuclear magnetic resonance (NMR) \cite{Cui2021} in film samples and  using neutron scattering \cite{Puphal2022} in bulk LaNiO$_2$, where H appears to cluster at the grain boundaries.

The work by Ding {\em et al.}~\cite{Ding_Nature_2023_NdNiO2_H}
now prompts for a complete overhaul of our picture of superconductivity in nickelates.
Systematically increasing the exposure time to CaH$_2$ and using ion mass spectroscopy,  Ding {\em et al.}
link the occurrence of superconductivity to a narrow range of hydrogen concentration $0.22<x<0.28$. Most notably, superconductivity seems absent for
low hydrogen concentrations, implying that its presence
is necessary for superconductivity. Arguably the most obvious --but hitherto for nickelates unexplored-- route for hydrogen to cause superconductivity is
via the conventional, electron-phonon (\emph{ep}) mechanism.
Due to its light mass, hydrogen can lead to high-temperature superconductivity, with critical temperatures ($T_c$) up to almost room temperature in hydrides under pressure~\cite{Eremets_Nature_2015_SH3, Eremets_NatPhys_2016_SH3, Hemley_PRL_2019_LaH, Boeri_JPCM_2019_viewpoint, Pickard_AnnRevCMP_2020_review}. Furthermore, the $s$-wave-gap reported in Ref.~\cite{PhysRevLett.129.077002} might be naturally explained
from such an \emph{ep} mechanism.
In this context, a hydrogen \ep{} mechanism for at least some of the superconductivity in nickelates appears to be a very reasonable and appealing working hypothesis. 
Let us also note that the \emph{ep} mechanism for nickelate superconductors without hydrogen has been explored previously, but does not result in sizable $T_c$'s within density-functional theory (DFT) \cite{Nomura2019}.

\begin{figure}[tb]
	\includegraphics[width=0.95\columnwidth]{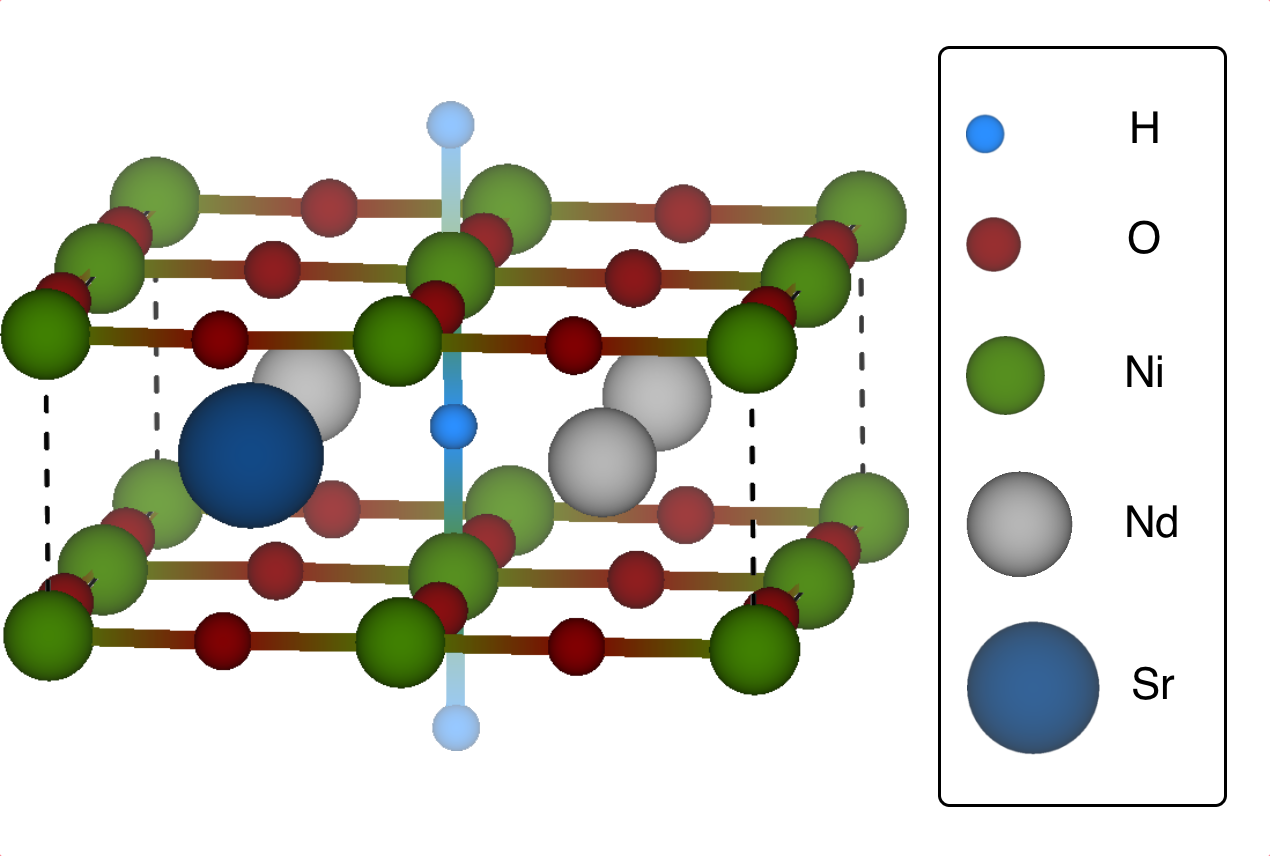}
	\caption{2$\times$2$\times$1 supercell for Nd$_{0.75}$Sr$_{0.25}$NiO$_{2}$H$_{0.25}$ with the experimentally optimal hydrogen concentration~\cite{Ding_Nature_2023_NdNiO2_H}. The hydrogen chain is indicated by two additional
        H atoms outside the supercell;  atoms at the surface, edge, and corner count by a factor of 1/2, 1/4 and 1/8 respectively.}
	\label{fig:figure1}
\end{figure}

In this letter, we thus explore the hydrogen  \emph{ep} scenario for
Nd$_{0.75}$Sr$_{0.25}$NiO$_2$H$_{0.25}$ which is exactly at the optimum of Ding {\em et al.}~\cite{Ding_Nature_2023_NdNiO2_H}.
Since it is energetically favorable for hydrogen to form chains
\cite{Si_PRL_2020_LaNiO2_H, Si_arXiv_2022_LaNiO2_H}, the simplest structure
compatible with the experimental observation
is one hydrogen (chain) in a
2$\times$2$\times$1 supercell, see Fig.~\ref{fig:figure1}.
This supercell can also accommodate 25\% Sr doping, close to the experimentally
investigated 20\% and still in the range of superconducting dome \cite{li2019superconductivity,Zeng2021,lee2022character}.
We investigated the electronic and vibrational properties  by means of DFT \cite{PhysRev.136.B864} and density-functional perturbation theory (DFPT) \cite{RevModPhys.73.515}, respectively. We find however that the electron-phonon (\emph{ep}) coupling is minimal, and cannot explain the reported $T_c$'s. Furthermore, engineering optimal conditions for \emph{ep} superconductivity by changing the rare earth to La and performing a comprehensive study of different hydrogen concentrations does not yield any finite transition temperature either. We thus conclude that the measured $T_c$ eludes an explanation in terms of a simple boost in the \emph{ep} coupling driven by hydrogen.

{\em Methods: DFT.}
All DFT calculations were performed using \verb|Quantum ESPRESSO| version 7.1, employing optimized norm-conserving Vanderbilt pseudopotentials \cite{Hamann_PRB_2017_ONCV, vanSetten_CompPhysComm_2018_PSDOJO}. The pseudopotential of neodymium uses the frozen-core approximation for the $f$ states. We used a 90~Ry cutoff on the plane-waves expansion, and 8$\times$8$\times$8 grid with a 0.040~Ry smearing for Brillouin zone integration. The crystal structures were constructed using VESTA \cite{Izumi_JAC_2008_VESTA} and subsequently relaxed until forces (stresses) were lower than 10$^{-5}\,$Ry/Bohr (0.5 kBar). Due to the larger size of Sr compared to Nd a local distortion of the Ni-O-Ni bond angle is induced, which deviates from 180 to 172$^{o}$ around Sr to accommodate the atom; see Supplementary Material~\cite{suppmat} Fig. S1.

Phonon calculations were performed on a $\Gamma$-centered 2$\times$2$\times$2 grid, within the harmonic approximation. Anharmonic corrections were introduced for specific modes using the frozen-phonon approach presented in \cite{Heil_PRL_2017_NbS2} (further details are available in the Supplemental Material~\cite{suppmat}). The integral of the electron-phonon matrix elements was performed on a 16$\times$16$\times$16 and 24$\times$24$\times$24  grid. A Gaussian smearing of 100~meV was found to give a converged result. The rigid-band approximation for the integration of the \ep{} matrix elements was performed using our modified version of \verb|Quantum ESPRESSO|\cite{our_qe_footnote}.

{\em Methods: McMillan formula.}
In a conventional \ep{} superconductor the superconducting \Tc{} can be estimated by the McMillan formula \cite{McMillan_PR_1968_Tc, Allen_PRB_1975_McMillan}, which works particularly well in the weak-coupling regime:
    \begin{equation}
    \label{eq:mcmillan}
      T_{c} = \frac{\omega_{log}}{1.2} \exp{\left[ \frac{1.04(1+\lambda)}{\lambda-\mu^{*}(1+0.62\lambda)} \right]}
    \end{equation}
where $\lambda$ and $\omega_{log}$ describe the average strength of the \ep{} coupling and phonon energies, respectively. It is apparent from Eq. \ref{eq:mcmillan} that high \Tc{}'s require a combination of both (i) strong \ep{} coupling (large $\lambda$) and (ii) high phonon energy (large $\omega_{log}$). Generally speaking, hydrogen can boost both of these quantities as phonons involving it are typically high-energy and unscreened; and hydrogen-rich conventional superconductors have indeed reached extremely high \Tc{}'s \cite{Eremets_Nature_2015_SH3, Eremets_NatPhys_2016_SH3, Hemley_PRL_2019_LaH, Boeri_JPCM_2019_viewpoint, Pickard_AnnRevCMP_2020_review}.

Fermi surface nesting for Nd$_{0.75}$Sr$_{0.25}$NiO$_2$H$_{0.25}$ was computed using the  \verb|EPW| code~\cite{Giustino_PRB_2007_epw, Giustino_CPC_2016_EPW}, with a Wannier interpolation over a 32$\times$32$\times$32 grid (for further details on the wannierization we refer the reader to the Supplemental Material \cite{suppmat}). In Nd$_{0.75}$Sr$_{0.25}$NiO$_2$H$_{0.25}$ \verb|EPW| was also used to provide an additional convergence test of \ep{} properties by interpolating the \ep{} matrix elements over a 12$\times$12$\times$12 and 24$\times$24$\times$24 grid for phonons and electrons, respectively, still yielding a \Tc{} of 0~K.

\begin{figure}[tb]
	\includegraphics[width=0.99\columnwidth]{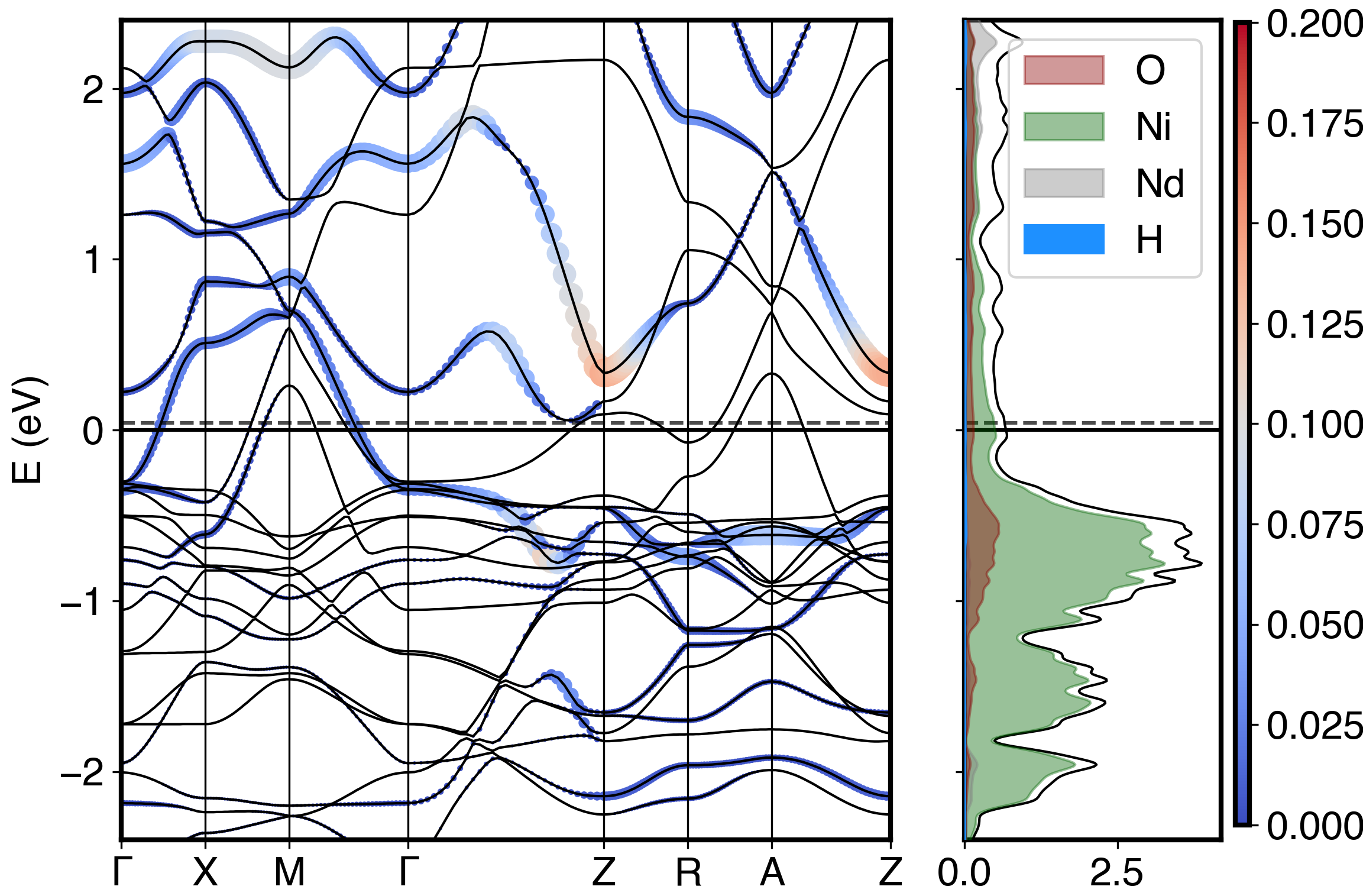}
	\caption{left: Electronic band structure of 
Nd$_{0.75}$Sr$_{0.25}$NiO$_2$H$_{0.25}$, decorated with the orbital projection onto the H-$1s$ state. Right: Total and atom-projected density of states. The size and color scale of the colored bands (left) indicates the H fraction of states, from 0 to 0.20. The Fermi energy for 25\% and 20\% Sr doping is shown as a solid black and dashed gray line, respectively. The DOS (right) is in units of states/eV/f.u. The atom projection onto Nd(+Sr), Ni, O, and H is shown as gray, green, red, and blue-filled curves, respectively.}
	\label{fig:figure2}
\end{figure}

{\em Electronic structure.}
In Fig.~\ref{fig:figure2} we show the electronic band structure (left) decorated with the hydrogen $1s$ character, and the orbital resolved density of states (right) of Nd$_{0.75}$Sr$_{0.25}$NiO$_{2}$H$_{0.25}$ (a comparison with the band structure of the parent compound Nd$_{0.75}$Sr$_{0.25}$NiO$_{2}$ is shown in the Supplemental Material \cite{suppmat} Figure S2;  it has also been calculated before e.g.\ in~\cite{Wu2019,Nomura2019,Zhang2019,jiang2019electronic}). The presence of topotactic hydrogen opens a wide gap around the $Z$ point, slightly above the Fermi energy, and another one in the band going from $\Gamma$ to $Z$. These bands can be identified easily since they present a significant hydrogen character. Due to its low concentration relative to the other elements, hydrogen contributes to only about 2\% of the DOS at the Fermi level. A band with significantly higher hydrogen character is present along the $\Gamma$-$Z$ direction, at about 0.5~eV above the Fermi energy, which is, however, too high to contribute significantly to superconductivity. 

The Fermi surface consists of three sheets (shown in Supplemental Material \cite{suppmat} Fig. S3): a long, tubular sheet forming an electron pocket around $\Gamma$, and similar hole pockets around $M$, elongated along the $k_{z}$ direction. None of these sheets presents significant hydrogen character, which is rather evenly spread over all wavevectors.

\begin{figure}[tb]
	\includegraphics[width=\columnwidth]{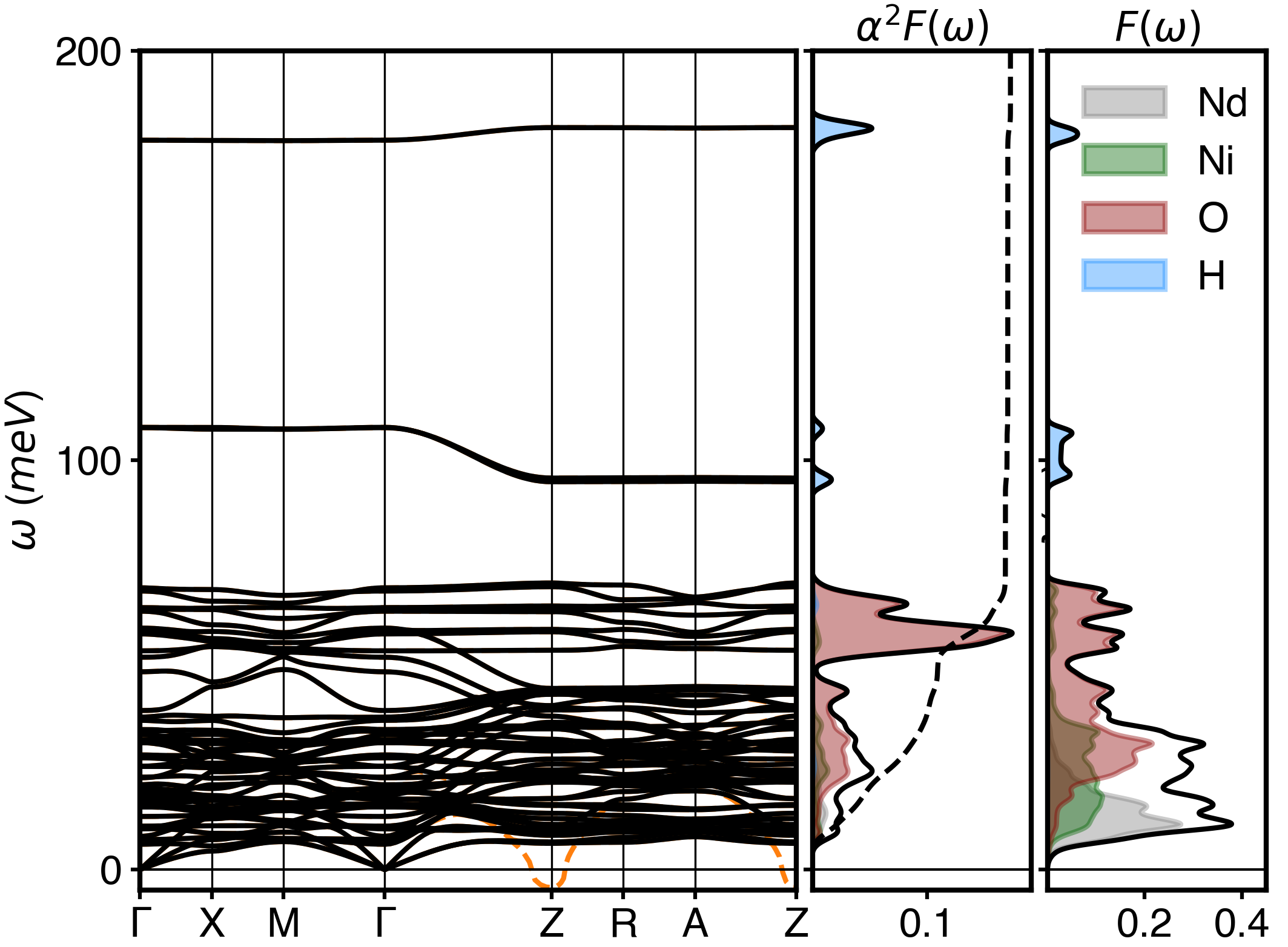}
	\caption{Phonon dispersions (left), atom-projected Eliashberg function ($\alpha^2F(\omega)$; middle), and phonon density of states ($F(\omega)$; right). In the harmonic approximation, there is an instability
          at the $Z$ point (orange dashed orange lines), which is removed when including the anharmonic correction at this point
          (black solid lines). The total phonon DOS and Eliashberg function are shown as black solid lines, while projections onto Nd(+Sr), Ni, O, and H are shown as gray-, green-, red-, and blue-filled curves.}
	\label{fig:figure3}
\end{figure}

{\em Electron-phonon superconductivity.}
To establish whether the presence of hydrogen leads to a significant superconducting \Tc{} via the conventional \ep{} mechanism, we computed the superconducting properties using DFPT as implemented in \verb|Quantum ESPRESSO| \cite{quantumespresso_1, quantumespresso_2, Baroni_RevModPhys_2001_DFPT}. In Fig. \ref{fig:figure3} we report the phonon dispersions along with the atom-projected phonon density of states and the Eliashberg function. 

The phonon dispersion is characterized by two rather flat branches at about 110 and 180~meV, which correspond to the twofold-degenerate in-plane (Nd-H) and out-of-plane (Ni-H) hydrogen vibrations. In addition, a single mode involving in-plane bending of the Ni-O bond presents a small imaginary frequency at the $Z$ point. Inclusion of anharmonic effects via a frozen-phonon approach as in~\cite{Heil_PRL_2017_NbS2} is enough to remove this instability, as the anharmonic mode goes from 6$i$ to 9 meV~\cite{imaginary_freq_footnote}. Using the same approach, we computed the anharmonic vibrational frequency for the Ni-H and Nd-H modes at the $\Gamma$ point. In the Ni-H mode, we observe a hardening of the mode from 178 to 191~meV, while for the Nd-H mode, we found the anharmonic frequency at 113 meV is only 5 meV higher than the harmonic result. To check the possible influence of anharmonicity on the \ep{} coupling, we diagonalized the dynamical matrix and computed the coupling both with and without it, but found no significant change in our results.


The phonon modes involving Ni-H and Nd-H exhibit only an extremely small \ep{} coupling and thus essentially do not contribute to superconductivity. Indeed, the integrated \ep{} coupling and average phonon frequency are $\lambda = 0.16$ and $\omega_{log} = 43.4\,$meV, respectively. This means that the superconducting \Tc{} estimated via the McMillan formula is essentially zero \cite{McMillan_PR_1968_Tc, Allen_PRB_1975_McMillan} and the \Tc{} observed in Ref.~\cite{Ding_Nature_2023_NdNiO2_H} cannot be explained. To further rule out the unlikely event that the small difference in doping between the experimental compound (Nd$_{0.8}$Sr$_{0.2}$NiO$_{2}$H$_{0.25}$) and our calculations (Nd$_{0.75}$Sr$_{0.25}$NiO$_{2}$H$_{0.25}$) induces a significant change in \Tc{}, we performed the same calculations for an effective 20\% Sr doping \cite{our_qe_footnote} using a rigid-band approximation. The results are summarized in Tab. \ref{tab:supercond_properties}; both $\lambda$ and $\omega_{log}$ remain essentially identical and the resulting \Tc{} is also zero. 

{\em Engineering optimal conditions for \ep{} superconductivity in nickelates.}
In the previous section we discussed the absence of \ep{} mediated superconductivity in Nd$_{0.75}$Sr$_{0.25}$NiO$_{2}$H$_{0.25}$. However, as previously noted, a stronger hydrogen character is present at about 0.5~eV above the Fermi level (Fig.~\ref{fig:figure2}), which might move towards the Fermi energy given a slight modification in the crystal structure and/or a higher electron filling. Since in superconducting hydrides the hydrogen character of states at the Fermi energy typically correlates with higher \Tc{}'s \cite{Errea_NatComm_2021_hydridebased, Boeri_JPCM_2021_roadmap}, this level of filling would appear more promising for the scenario of conventional superconductivity.

To explore this possibility and the effect of different rare earth cations, we studied different hydrogen configurations of the closely-related LaNiO$_2$H$_{x}$ compound. Indeed, a hydrogen concentration of 25\%, consistent with that reported in \cite{Ding_Nature_2023_NdNiO2_H}, yields an electronic structure that favors the previously outlined scenario. That is the bands with the largest hydrogen character that are 0.5~eV above the Fermi level for  NdNiO$_2$H$_{0.25}$ are crossing the Fermi energy for  LaNiO$_2$H$_{0.25}$ (see Supplemental Material \cite{suppmat} Fig. S5).
We thus computed the vibrational and superconducting properties for four different topotactic hydrogen concentrations $x= 11\;$\%, 22\%, and 55\% (in a 3$\times$3$\times$1 supercell), and 25\% (in a 2$\times$2$\times$1 supercell) for  LaNiO$_2$H$_{x}$. A summary of these results is shown in Tab. \ref{tab:supercond_properties}. Despite the more favorable conditions for \ep{} mediated superconductivity,  we find a total \ep{} coefficient $\lambda$ no higher than 0.21 in all the configurations investigated. Thereby we confirm that, albeit the contribution of hydrogen can be slightly more significant, the \ep{} coupling remains low and cannot explain the observed \Tc{}'s. 

\begin{table}[tb]
    \centering
    \begin{tabular}{|c|c|c|c|}
    \hline
    Composition     & \quad $\lambda$ \quad \quad & \quad $\omega_{log}$~(meV) \quad   & \quad \Tc{}$^{**}$~(K) \quad \\
    \hline
    \hline
Nd$_{0.75}$Sr$_{0.25}$NiO$_2$H$_{0.25}$  &     0.16      &     43.4     &   0    \\
Nd$_{0.80}$Sr$_{0.20}$NiO$_2$H$_{0.25}$$^{*}$  &     0.17      &     44.1     &   0    \\
LaNiO$_2$H$_{0.11}$                   &     0.21      &     33.0     &   0    \\
LaNiO$_2$H$_{0.22}$                   &     0.21      &     36.5     &   0    \\
LaNiO$_2$H$_{0.25}$                   &     0.21      &     42.8     &   0    \\
LaNiO$_2$H$_{0.55}$                   &     0.17      &     39.0     &   0    \\
    \hline
    \end{tabular}
    \caption{Summary of the calculated superconducting properties of various nickelate compounds with topotactic hydrogen. $^*$: calculated by shifting the Fermi energy. $^{**}$: calculated \Tc{}'s below 1 mK are considered as zero.}
    \label{tab:supercond_properties}
\end{table}

Our direct calculations of the \ep{} coupling thus show that the Ni-H or Nd/La-H bonds do not contribute sufficiently to the \ep{} coupling to explain the \Tc{}'s measured by~\cite{Ding_Nature_2023_NdNiO2_H}.
In NdNiO$_2$H$_{0.25}$, the \ep{} mechanism is not particularly supported by the electronic structure, since no bands that cross the Fermi energy exhibit significant hydrogen character -- an important ingredient for superconductivity in hydrides \cite{Heil_PRB_2018_noscFeH, Errea_NatComm_2021_hydridebased}. However, even when the hydrogen bands are located  at the Fermi energy they do not cause a significant \Tc{}. This is most likely due to the ionic character of the La-H and Ni-H bonds \cite{Heil_PRB_2015_bondingSH3}, which cause the \ep{} matrix elements to be small.

{\em Enhanced \Tc{} from Fermi surface nesting.}
Having established that the electron-phonon matrix elements are in general small, the only other scenario supporting conventional superconductivity could come from an enhancement due to Fermi surface nesting \cite{Kohn_PRL_1959_KohnAnomaly, Heil_PRL_2017_NbS2}. On a qualitative level, the square-like sheet of the Fermi surface could indeed support this for phonons with wavevectors $\vec{q}_{nest} \sim (0.0, 0.4, 0.0)$ and $(0.4, 0.0, 0.0)$
\cite{noteRIXS}.

To examine this possibility, we computed the Fermi surface nesting function, defined as in Refs.~\cite{Heil_PRB_2014_acc_baresus, Giustino_CPC_2016_EPW}, along a high-symmetry path (see Supplemental Material~\cite{suppmat} Fig. S4), which presents two local maxima at $X$ and $M$. Since these points were already present in the mesh used for the \ep{} calculations, we also rule out nesting as a possible source of elusive \ep{} interaction.

{\em Conclusion.}
We investigated the possibility of topotactic hydrogen inducing superconductivity in  nickelates 
through the conventional 
electron-phonon mechanism.
Experimentally  Sr-doped nickelates with a \Tc{} of about 15~K appear to be extremely sensitive to the hydrogen concentration and some experiments suggest an $s$-wave gap that is to be expected in this scenario. Notwithstanding, we find that hydrogen does not strongly affect the states at the Fermi surface and that the \ep{} coupling is too weak. The \ep{} mediated \Tc{} of 
Nd$_{0.75}$Sr$_{0.25}$NiO$_{2}$H$_{0.25}$ is thus essentially zero.
 
To rule out that we missed the optimal conditions for
 \ep{} mediated superconductivity, we
 further engineered the band structure by changing the rare-earth atom and hydrogen concentration. This did not yield any finite \Tc{} either.
 Given the very weak \ep{} coupling with $T_c<1\,$mK even under optimal conditions, we do not expect that many-body effects \cite{Si2019,Petocchi2020,PhysRevB.102.161118,PhysRevB.106.115132,Pascut2023} beyond our DF(P)T calculation such as quasi-particle renormalization, Hund's exchange on Ni, and modifications of crystal field splittings can enhance the \Tc{} significantly.
 Consequently, we are inclined to conclude that hydrogen-derived phonons do not mediate superconductivity in infinite-layer nickelate superconductors.
{\em Alternative explanations.}
So why
does a narrow range of hydrogen concentration appear to be essential for superconductivity in nickelates? One possibility is that (i)~the inclusion of hydrogen changes the electronic structure and environment in a manner that is favorable for a mechanism different from conventional \ep{} coupling. However, currently, none of the proposed mechanisms for superconductivity in infinite-layer nickelates relies on the presence of hydrogen. On the contrary, it has been argued \cite{Si2019,Jiang2019} that the spin-1 state and
three-dimensionality that is induced by topotactic hydrogen is unfavorable for superconductivity.
Furthermore, it is at least somewhat unexpected that the window of hydrogen, where superconductivity is found, appears fairly small. With its agility, hydrogen will also tend to spread through the crystal and thus induce some disorder which is generally unfavorable for superconductivity. 

Another possibility (ii) is that not only the hydrogen concentration is changed during the reduction process. Specifically, Ding {\em et al.}~\cite{Ding_Nature_2023_NdNiO2_H} use longer CaH$_2$ exposure times as a means to control the amount of intercalated H; and it is not unreasonable to surmise other aspects of the sample might alter as well. The most important is that these longer reduction times also affect the oxygen content which has to be reduced in the first place and which was not analyzed in Ref.~\cite{Ding_Nature_2023_NdNiO2_H}. The purported narrow range of hydrogen concentration might thus simply  be the sweet spot of reduction time in Nd$_{0.8}$Sr$_{0.2}$NiO$_{2+\delta}$H$_{x}$ with $\delta$ already sufficient  low but $x$ not yet too high for superconductivity --- and $\delta=x=0$ being the unreachable optimum.


{\em Acknowledgments}
S.D.C. thanks Lilia Boeri for  useful discussion, and Christoph Heil and Roman Lucrezi for sharing their code for calculating the anharmonic dynamical matrices. We acknowledge funding through the Austrian Science Funds (FWF) projects id I 5398, P 36213, SFB Q-M\&S (FWF project ID F86), and Research Unit QUAST by the
Deuschte Foschungsgemeinschaft (DFG; project ID FOR5249) and FWF
(project ID I 5868). L.S. is thankful for the starting funds from
Northwest University. Calculations have been done in part on the Vienna
Scientific Cluster (VSC).

{\em Data availability} 
Raw data and our modifications to \verb|Quantum ESPRESSO| are available at \url{XXX}.

\begin{thebibliography}{69}
\bibitem [{ima()}]{imaginary_freq_footnote}%
	\BibitemOpen
	\href@noop {} {}\bibinfo {note} {The imaginary frequencies are plotted as
		negative in Fig. \ref{fig:figure3}.}\BibitemShut {Stop}%
\end{thebibliography}
%

\end{document}